\documentclass{an}
\usepackage{graphicx}
\usepackage{times}
\usepackage{fancyhdr}
\begin{document}

\title{Close binary systems among very low mass stars and brown dwarfs}

\author{R. D. Jeffries and P. F. L. Maxted}
\institute{Astrophysics Group, Keele University, Keele, Staffordshire
  ST5 5BG, UK}

\date{Received; accepted; published online}

\abstract{
Using Monte Carlo simulations and published radial velocity surveys we
have constrained the frequency and separation ($a$) distribution of
very low mass star (VLM) and brown dwarf (BD) binary systems. We find
that simple Gaussian extensions of the observed wide binary
distribution, with a peak at 4\,au and $0.6<\sigma_{\log(a/{\rm
    au})}<1.0$,  correctly reproduce the
observed number of close binary systems, implying a close ($a<2.6$\,au)
binary frequency of 17--30 per cent and overall frequency of 32--45 per
cent. N-body models of the dynamical decay of unstable protostellar
multiple systems are excluded with high confidence because they do not
produce enough close binary VLMs/BDs. The large number of close binaries and
high overall binary frequency are also completely inconsistent with
published smoothed particle hydrodynamical modelling and argue against a
dynamical origin for VLMs/BDs.
}

\correspondence{rdj@astro.keele.ac.uk}

\maketitle

\section{Introduction}

Very low-mass stars (VLMs, $M<0.15\,M_{\odot}$) and brown dwarfs
(BDs, $M<0.075\,M_{\odot}$) are now known to be common in our galaxy, 
possibly even outnumbering stars of higher mass. Yet their origins are
still mysterious because the typical Jeans mass in a giant molecular
cloud is an order of magnitude more massive.

Recent ideas, suggest that ``turbulent'' fragmentation might allow
enhancements in local densities that can result in the formation of
{\it very} small cores (e.g. Padoan \& Nordlund 2004). However, the
biggest problem for the formation of VLMs and BDs may be how to prevent
these very small cores subsequently accreting too much and
becoming higher mass stars (Boss 2002). One solution is if VLMs and BDs
were originally part of unstable multiple systems and, as the least
massive components, are ejected on timescales too short for significant
mass accretion (Boss 2001; Reipurth \& Clarke 2001).

These ideas have proved very persuasive and it is possible that
observations of accretion disks or VLM/BD kinematics and spatial
distributions (see other contributions in this volume) may ultimately
provide crucial observational tests. A further observational constraint
is offered by the frequency and separation distribution of binary
systems and there have been many investigations along these lines in the
last few years.

Binary systems should preserve information about their formation,
particularly their dynamical origins. A plethora of studies have been
published, which use high spatial resolution imaging, either Hubble
Space Telescope or adaptive optics in the near-infrared, to find the
frequency of binary systems among nearby field VLMS/BDs which have
separations $a\geq 1$\,au (e.g. Bouy et al. 2003; Burgasser et al. 2003;
Close et al. 2003; Siegler et al. 2005).  The lower cut-off to the
separation range that can be probed is of course a result of
limited angular resolution.  The results of Close et al. are quoted
below, but all of these studies are in reasonable agreement. The main
findings are that the binary frequency for separations $a>2.6$\,au is
$15\pm 7$ per cent (see Fig.~1).  This frequency is lower than found
for G to M-type field stars (Duquennoy \& Mayor 1991; Fischer \& Marcy
1992), but these earlier studies counted binaries over a much wider range of
separations, including close, spectroscopic binary systems.

Of more importance may be that all these studies failed to find binary
VLMs/BDs with $a>15$\,au. This has been seized upon as supporting
evidence for the ``ejection'' scenario where qualitatively, one expects
fewer wide binary systems with low binding energies to survive the
ejection process. There are several problems with this: (i) some wider
VLM/BD binary systems are starting to be found (e.g. Luhman 2004;
Phan-Bao et al. 2005); (ii) some theoretical models that produce BDs by
what could be termed ``ejection'' are also producing some BD-BD
binaries with wide separations (Bate \& Bonnell 2005); (iii) there may
yet be alternative means of breaking up wide binaries during their
first few Myr in a dense cluster environment and then one would not
expect to see them in the field or older clusters.

The frequency of {\it close} ($a<2.6$\,au) binary systems may provide
more secure evidence, but little progress has so far been made.
Guenther \& Wuchterl (2003) found 3 radial velocity (RV) variables
among 24 field VLMs/BDs; Joergens (2005) found 2 RV variables from a
sample of 11 VLMs/BDs in the Chamaeleon I star forming region; and
Kenyon et al. (2005) found evidence for 4 close binaries in a sample of
$\sim 60$ VLMs/BDs in the young Sigma Ori cluster.  Taken together,
these results {\it hint} at a surprisingly high ($>10$ per cent)
frequency for close VLM/BD binaries.  In this contribution we examine
this possibility with some statistical rigour and comment on the
implications of our results for the formation mechanism of VLMs/BDs.

\begin{figure}
\resizebox{\hsize}{!}
{\includegraphics[]{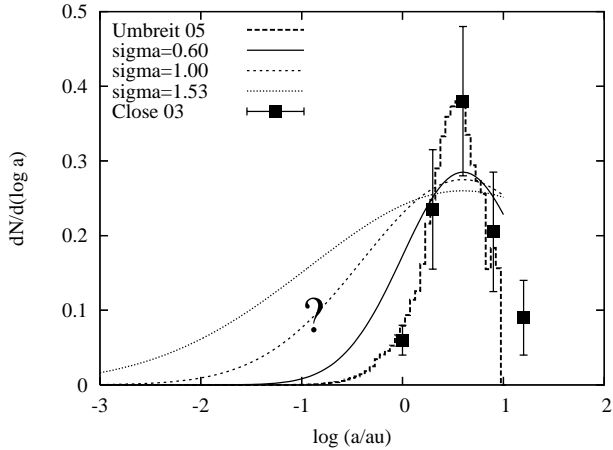}}
\caption{The observed distribution of VLM/BD binaries as a function of
  $\log a$ (from Close et al. 2003). Note that the point at $\log
  (a/{\rm au})=0$ is significantly affected by incompleteness.  Also
  shown are the various model $\log a$ distributions used to simulate
  the VLM/BD binary population (see text). They are from Umbreit et
  al. (2005) or are Gaussian distributions with peaks at 4\,au and with
  various indicated widths (truncated for $a>10$\,au).  The model
  distributions are normalised such that the binary frequency is 15 per
  cent for $2.6<a<10\,$au.
}
\end{figure}

\section{Analysis}

We have collected RV measurements for VLMs/BDs from three main sources
in the literature.  Consideration is restricted to those objects with
RV measurements at $\geq 2$ epochs. There are 24 field objects with
$0.06<M/M_{\odot}<0.10$ from Guenther \& Wuchterl (2003), 10 objects
with $0.05<M/M_{\odot}<0.10$ from Joergens (2005) and 14 objects with
$0.045<M/M_{\odot}<0.11$ from Kenyon et al. (2005).  Masses were
either estimated from spectral types (assuming an age of 1\,Gyr for
field stars) or from isochrones in colour magnitude diagrams.

We identified possible binary systems by calculating the value of
$\chi^{2}$ for a constant value fitted to the RV measurements of each
object. Potential binaries are those for which the $\chi^{2}$ value
corresponds to a probability, $p$, of a constant RV that is less than
$10^{-3}$. This threshold was chosen to avoid contaminating what turns
out to be a low number of binaries with any spurious statistical detections. 
During the course of this procedure we found, by
investigating the distribution of $p$ values, that the RV uncertainties
were underestimated in the case of the Guenther \& Wuchterl (2003) and
Kenyon et al. (2005) data (for details see Maxted \& Jeffries 2005). 
Errors amounting to 0.4 and 4.5\,km\,s$^{-1}$ respectively
were added in quadrature to RV uncertainties from these two datasets.
We identified four objects as binaries: 2MASSWJ2113029-10094 and
LHS\,292 from Guenther \& Wuchterl, Cha\,H$\alpha$~8 from Joergens
(2005), and KJNOM~72 from Kenyon et al.

A Monte-Carlo simulation has been used to predict how many binary
systems we ought to have found, given the sensitivity and time sampling
of the data and for various assumptions about the frequency, and
distributions of $a$, eccentricity and mass ratio for binary
systems. The details are provided by Maxted \& Jeffries (2005) but
briefly, we have assumed various Gaussian extensions of the observed
$\log a$ distribution of VLM/BD binary systems, peaking at 4\,au and
with widths ranging from $\sigma_{\log (a/{\rm au})}=0.6$ to
$\sigma_{\log (a/{\rm au})}=1.53$ (corresponding to the width of the
field G-star distribution found by Duquennoy \& Mayor 1991). These
distributions were truncated at $a=10$\,au in order to better match
what is known about wider binary systems.  We also trialled the
prediction from a triple-body decay calculation for BD binaries from
Umbreit et al. (2005) (see Fig.~1).  The eccentricities were taken as
$e=0$ for binary periods less than 10~days and then randomly
distributed between $0<e<e_{\rm max}$ for longer periods, with $e_{\rm
max}=0.6$ or $e_{\rm max}=0.9$. The mass ratio distribution was assumed
either uniform for $0.2<q<1.0$ (the ``qflat'' distribution) or
uniform for $0.7<q<1.0$ (the ``qpeak'' distribution). These $q$
distributions seem reasonable compared with what is already known about
wider low-mass binary systems (e.g. see Bouy et al. 2003). It is
possible, that, as in more massive binary systems, there is a shift
towards equal masses among very close binary systems.  Simulated
targets were randomly assigned binary status on the basis of a total
binary frequency which was normalised such that the frequency was 15
per cent for $a>2.6$\,au and hence matched the observational
constraints for wider binaries.

\section{Results and Discussion}

\begin{figure}
\resizebox{\hsize}{!}
{\includegraphics[]{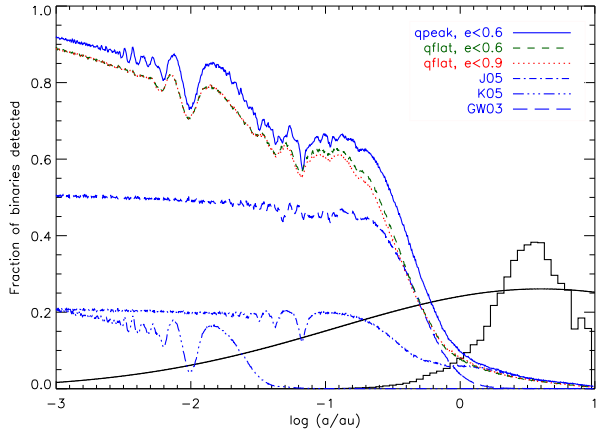}}
\caption{The average binary detection efficiency. The upper three
  curves show the effects of differing models for the mass ratio and
  eccentricity distributions (see text). The lower curves show the
  contributions to the average from the each of the three RV surveys
  considered (i.e. the sum of these gives the average). 
  To illustrate the binary population we might be sensitive
  to, we show the two extreme model $a$ distributions we have
  investigated (see Fig.~1).
}
\end{figure}

The outputs from our modelling are a prediction of the sensitivity of
each of the target observations (for each star) to binaries of a given
separation (averaged over all mass ratios and eccentricities).  We plot
the average sensitivity for all targets in Fig.~2 and also show how the
data from each of the three datasets contributes to this average
(i.e. the lower curves are the sum of the detection efficiencies for
each target in that datsaset, divided by the number of targets in all
three datasets).  It is clear from this that we are sensitive to
binaries with $a\leq 1$\,au. It is also clear that we cannot simply say
binaries are all detectable out to some limiting separation and cannot
be detected beyond that. Even as a function of $\log a$, the detection
efficiency rolls over rather gradually, and combined with the
(probable) increasing number of binary systems as $\log a$ increases,
this demonstrates why a detailed simulation is demanded.  From Fig.~2
we can see that narrower $a$ distributions such as that of Umbreit et
al. (2005) predict far fewer binaries will lie within our sensitive
range.  This is further illustrated in Fig.~3 where we show the
simulated distribution of $\log p$ values (where $p$ is the probability
that a set of RV measurements represent a constant value) for a million
binary systems with orbital separations taken from the various
models. The number of detected binary systems (with $\log p < -3$) is
expected to vary by an order of magnitude depending upon the adopted
$a$ distribution.

\begin{figure}
\resizebox{\hsize}{!}
{\includegraphics[]{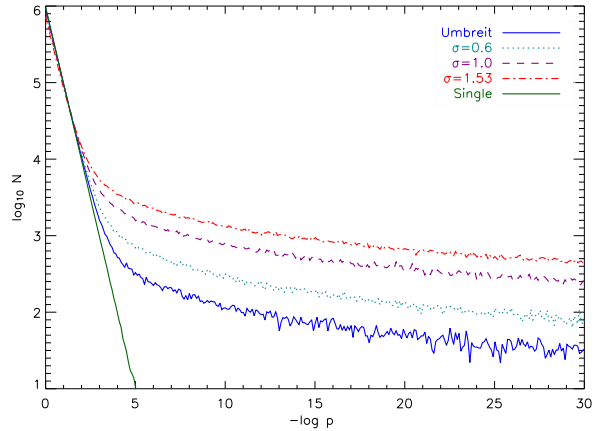}}
\caption{The distribution of $\log p$ (averaged over all the objects
  included in the RV surveys and where $p$ is the probability of the
  measured RVs being consistent with a constant value) for a simulated
  population of $10^{6}$ binaries. Separate curves are shown for 
  binary separations which were drawn from each of the considered model
  $a$ distributions. The straight line is the predicted locus for
  single stars.
}
\end{figure}

\begin{table}[h]
\caption{The number of binaries, $n_{\rm bin}$, predicted from the Monte-Carlo
  simulations. The total binary fraction is $f_{b}$ and probabilities
  (as percentages)
  are quoted for whether the simulations would result in 4 or more, or 4 or fewer
  detected binaries in the observational dataset.}
\begin{tabular}{ccccc}\hline
$F (\log a)$  & $f_{b}$ & $n_{\rm bin}$ & $P(n_{\rm obs}\geq 4)$ &  $P(n_{\rm obs}\leq 4)$\\
\hline
Umbreit       & 0.26 & 0.5--0.7 & 0.1--0.4  & 99.9--99.9\\
$\sigma=0.60$ & 0.32 & 1.3--1.8 & 3.4--9.5  & 96.9--99.3\\
$\sigma=1.00$ & 0.45 & 4.2--5.5 & 62.9--82.5& 33.6--57.6\\
$\sigma=1.53$ & 0.59 & 9.0--10.8& 98.9--99.9& 0.8--3.3 \\
\hline
\end{tabular}
\end{table}

The key numerical results are given in Table~1, where we give the $a$
distribution assumed, the implied total binary frequency, the range for
the expected number of detected binary systems (summed over the three
datasets for various combinations of the $e$ and $q$ distributions) and
the probabilities that we would have observed either 4 or more, or 4 or
fewer binary systems in the real data. We find that the chosen $e$ or
$q$ distribution make little difference to the number of binary system
detections expected (as implied by the similar average detection
efficiency curves shown in Fig.~2). If binaries are really concentrated
towards lower $q$ values then we would expect to detect fewer of them
and so would require a higher binary frequency to explain the
data. Conversely, the minimum binary frequency would arise from a
population with $q=1$. There is no evidence that either of these
extremes are appropriate.  The upper end of the quoted range is
produced by the ``qpeak'' $q$ distribution. We {\it have} corrected the
expected number of binary systems for the well known bias in favour of
binary detection in magnitude limited samples of field stars (for
details see Maxted \& Jeffries 2005).

The expected number of detected binaries is very sensitive to the $a$
distribution, ranging from $<1$ for the Umbreit et al. (2005)
distribution to 11 for the $\sigma_{\log (a/{\rm au})}=1.53$
distribution.  The narrowest (Umbreit et al.)  $a$ distribution is
ruled out by the observations at the $>99$ per cent level. There are
too many binaries detected in the RV data.  The broadest ($\sigma_{\log
(a/{\rm au})}=1.53$) $a$ distribution is also highly unlikely for the
opposite reason.  The intermediate width distributions are in
satisfactory agreement with the observed number of close binaries and
imply overall binary frequencies of 32--45 per cent, with 17--30 per
cent at $a<2.6$\,au. This high close binary frequency is larger than
for either G-dwarfs (14 per cent, Duquennoy \& Mayor 1991) or M-dwarfs
($\simeq 10$ per cent, Fischer \& Marcy 1992). However, this is merely
a continuation of the trend which {\it is already apparent} in the
binaries with $2.6<a<10$\,au.

To clarify some points that might puzzle the reader, we note that: (i)
We do not {\it know} the correct functional form for the $a$
distribution, we have merely assumed a plausible form. To minimize the
implied binary frequency requires an $a$ distribution that is
concentrated towards small $a$ but which ``joins on'' to the observed
distribution at large separations. Of course one could hypothesize a
discontinuous distribution -- for example a secondary peak of binaries
at $a<0.1$\,au. Whatever distribution is assumed however, the binary
frequency for $a<2.6$\,au cannot be less than the 4/48 binaries we have
observed even if we were 100 per cent efficient at detecting binaries
out to that separation (which were are not by some margin -- see Fig.~2). 
(ii) The model $a$ distributions are normalised
to 15 per cent for $a>2.6$\,au, but this number is uncertain. If it
were increased to 22 per cent then tests show that narrower
distributions would be allowed (though the Umbreit et al. distribution
would still be ruled out at high significance), there would be slightly
fewer binary systems at $a<2.6$\,au, but the overall binary frequency
would be very similar. Conversely, a binary frequency of 8 per cent for
$a>2.6$\,au would require broader $a$ distributions to explain the
number of observed RV variables, more close binary systems, but again
the overall binary frequency would be almost the same. (iii) The 17--30
per cent figure for close binaries is not a formal error range - it is
the range of frequencies implied if the Gaussian forms assumed for the
$a$ distribution are correct and the observed binary frequency is
4/48. (iv) $q=1$ binaries may be harder to detect than we have
calculated. Our simulations assume the secondary star is ``invisible'',
but clearly an equal mass binary system will have an SB2 spectrum. For
close systems this will be obvious, but more separated systems, where
the orbital velocity amplitude is smaller than the spectral resolution
(certainly the case at $a\simeq 1$\,au for the datasets considered
here), would not immediately reveal themselves and might appear to have
constant RV. This effect can only act to {\it increase} the binary
frequency inferred from the observations.

Our results pose considerable difficulties for current ideas of VLM/BD
formation. When multiple systems form by fragmentation, the minimum
separation should be $a\ga 10$\,au due to the opacity limit.  Closer
binaries subsequently require the operation of dynamical or
hydrodynamical hardening mechanisms (Bate, Bonnell \& Bromm 2002).
N-body decay models of unstable multiple systems produce very few close
binary systems (e.g. Sterzik \& Durisen 2003; Umbreit et
al. 2005). This is because dynamical interactions capable of hardening
a VLM/BD binary usually result in the ejection of one of the components
and its replacement by a more massive object. Hydrodynamical hardening
process are of course not included in pure N-body models. The high
close binary frequency we have deduced indicates that these mechanisms
must be important and effective.

Smoothed particle hydrodynamic (SPH) models also have problems.  Bate
et al. (2002) and Bate \& Bonnell (2005) present simulations which
produce only $\sim 8$ per cent binarity among BDs. Whilst the rarity of
wide separation systems is approximately reproduced, the simulations also
yield very few binaries with $a<5$\,au. Even though the SPH models
are currently limited to resolutions of 1\,au, the lack of systems
produced with separations of a few au indicates that they should not be
common even when computational resolutions improve.  It appears to be a
common feature of both N-body and SPH models that the overall frequency
of VLM/BD binary systems formed through the decay of initially unstable
multiple systems is much lower than in higher mass stars
(Delgado-Donate et al. 2004; Hubber \& Whitworth 2005).

\section{Conclusions}

From the detection of 4 close VLM/BD binary systems in several
published RV surveys, we have been able to place strong constraints on
the binary frequency and separation distribution of such systems.  We
find that there are too many detected close binaries to be consistent
with the rather narrow separation distributions predicted by N-body
models of dynamically decaying multiple systems.  Instead we find that
broader Gaussian distributions with 17--30 per cent binarity for
$a<2.6$\,au are favoured and overall binary frequencies of 32--45 per
cent. The total binary frequency for VLMs/BDs is therefore only
marginally lower than (if not similar to) that for M-dwarfs, but
significantly lower than for G-dwarfs. Considering only separations
$a<10$\,au, the binary frequency of VLMs/BDs is higher than for either
G- or M-dwarfs.

The neglect of hydrodynamical hardening processes in N-body models may
be responsible for their lack of predicted close binaries. However,
recent SPH models also do not predict a significant population of very
close BD binaries and yield overall binary frequencies of only $\sim 8$
per cent.  The high overall binary frequency and the frequency of
observed close binary VLMs/BDs do not favour the ``ejection''
hypothesis or similar models that produce BDs from the decay of
dynamically unstable multiple systems. A mechanism is required that
hardens VLM/BD binaries without disrupting them.

At present, these conclusions are limited only by the small numbers of
detected binaries in the RV datasets. It is now important to increase
these samples with further high precision RV surveys of VLMs/BDs. Such
surveys will require sparse monitoring of large samples with baselines
of a year or more and RV precisions of $\simeq 1$\,km\,s$^{-1}$.  Not
only can the techniques we have demonstrated yield the binary
frequency, but modelling the observed $\log p$ distribution (as shown
in Fig.~3) offers a means of determining the actual form of the $a$
distribution at small separations, providing sufficient binaries can be
detected.


\end{document}